\documentclass[11pt,nofootinbib,article]{article}
\usepackage{epsfig}
\usepackage{amscd}
\usepackage{amssymb}
\usepackage{amsfonts}
\usepackage{amsmath}
\usepackage{pstricks}
\usepackage{pst-node}
\usepackage{mathrsfs}
\usepackage{verbatim}   
\usepackage{hyperref}
\usepackage{graphics,epsfig,psfrag}

\def\Tr{\mathrm{Tr}}
\def\tr{\mathrm{tr}}

\def\det{\mathrm{det}}
\def\ess{\kappa}

\def\cO{{\cal O}}
\def\cR{{\cal R}}

\def\cL{{\cal L}}

\def\bi{s}

\def\be{\begin{equation}}
\def\ee{\end{equation}}
\def\bea{\begin{eqnarray}}
\def\eea{\end{eqnarray}}
\def\eq#1{(\ref{#1})}

\begin{document}

\centerline{\bf \large Quantization and fixed points of non--integrable Weyl theory}
\bigskip
\centerline{C. Pagani and R. Percacci}
\medskip
\centerline{SISSA, via Bonomea 265, 34136 Trieste, Italy}
\centerline{and INFN, Sezione di Trieste, Italy}

\begin{abstract}
We consider a simple but generic model of gravity where 
Weyl--invariance is realized thanks to the presence of a gauge field for dilatations.
We quantize the theory by suitably defining renormalization group flows that describe the
integration of successive momentum shells, in such a way that
Weyl--invariance is maintained in the flow.
When the gauge fields are massless the theory has, in addition to Weyl invariance, 
an abelian gauge symmetry.
According to the definition of the cutoff, the flow can break or preserve
this extended symmetry.
We discuss the fixed points of these flows.
\end{abstract}

\section{Introduction}

Weyl's unified theory of gravitation and electromagnetism 
\cite{weyl} was the first modern example of a gauge theory.
\footnote{See \cite{oraifeartaigh} for a historical perspective.}
An attempt was made to identify local dilatations
\be
\label{wt}
g_{\mu\nu}(x)\rightarrow \Omega^2(x)g_{\mu\nu}(x)\ ,
\ee
nowadays called Weyl transformations, as the gauge transformations of electromagnetism.
It was noticed that the non-metric connection 
\be
\hat\Gamma_\mu{}^\lambda{}_\nu=
\Gamma_\mu{}^\lambda{}_\nu-
\delta^\lambda_\mu b_\nu-\delta^\lambda_\nu b_\mu
+g_{\mu\nu} b^\lambda\ ,
\ee
is invariant under \eqref{wt}, provided the vector field $b_\mu$
transforms as
\be
\label{gt}
b_\mu \to b_\mu+\Omega^{-1}\partial_\mu\Omega\ ,
\ee
which is formally identical to a gauge transformation of electromagnetism.
For any tensor $t$ of dimension $L^w$ one can then define a 
diffeomorphism-- and Weyl--covariant derivative $Dt$ by
\be
\label{wcovder}
D_\mu t=\hat\nabla_\mu t-w b_\mu t\ ,
\ee
where all indices have been suppressed.
With this covariant derivative, and its curvature tensor $\cR$ defined by
\be
\label{wcurv}
[D_\mu,D_\nu]v^\rho=\cR_{\mu\nu}{}^\rho{}_\sigma v^\sigma\ .
\ee
it is then easy to construct diffeomorphism-- and Weyl--invariant actions.

This theory was immediately criticized by Einstein and soon fell in disfavor.
\footnote{A later attempt to revive it by Dirac \cite{dirac} 
is also flawed \cite{bekenstein}.}
We now understand that electromagnetism is a gauge theory of a compact abelian
gauge group acting on complex quantum mechanical wave functions,
rather than a non-compact abelian group of dilatations.
Still, Weyl's theory remains physically viable, provided we do not insist
on identifying $b_\mu$ with the electromagnetic potential.
Rather, one has to think of it as a component of the gravitational connection $\hat\Gamma$.
Weyl's theory can be seen as one of the simplest 
examples of gravitational theories with independent metric and connection:
in this case the torsion is zero, but the connection is not metric.
In this guise, the field $b_\mu$ is often used in conformal supergravities.
It is much less studied in the context of non-supersymmetric theories of gravity.
It is an interesting question to put bounds on the possible strength
of such non-metric parts of the gravitational connection, but we shall refrain
from discussing this issue in this paper.
Instead, we shall be interested in the quantum properties of the field $b_\mu$,
and in the fate of Weyl invariance under quantization.

Weyl invariance is a local version of scale invariance, and scale invariance
is generally broken when the theory is quantized.
This happens because in the process of field quantization one always has to
introduce some dimensionful parameter (cutoff, renormalization scale etc.).
This is the phenomenon of the anomaly, which generically manifests itself as a nonvanishing
trace of the expectation value of the energy-momentum tensor \cite{duff}.
One is used to the inevitability of this phenomenon.
On the other hand, when scale invariance is gauged by the introduction of the
gauge potential $b_\mu$, things look very much as when one gauges an internal
group and since the existence of gauge anomalies is not a generic phenomenon
one should not be too surprised if Weyl invariance could be preserved under quantization.
In this paper we will discuss an example of Weyl--invariant 
quantization of a classically Weyl--invariant theory.

This had been discussed earlier in the case when the field $b_\mu$ is a pure gauge $b_\mu = -\chi^{-1}\partial_\mu\chi$ \cite{percacci,cdpp}.
Then one can use the scalar field $\chi$ (which we call dilaton)
as a compensator which absorbs all Weyl--non--invariance,
both in the classical and in the quantum theory.
It was shown in \cite{cdpp} that even though Weyl invariance is preserved,
the trace anomaly is still present with all its physical consequences.
In this paper we will treat the case when $b_\mu$ is not a pure gauge.

We consider the most general class of Weyl-invariant actions for 
$g_{\mu\nu}$, $b_\mu$ and $\chi$ that contain at most two derivatives, see equation (\ref{action1}).
It defines a four-dimensional theory space.
In this theory Weyl invariance is ``higgsed'': in the ``unitary'' gauge the
kinetic term of $\chi$ becomes a mass term for $b_\mu$.
However, there is a three-dimensional subspace of theories where $b_\mu$ is massless
and an additional abelian gauge invariance appears.
In addition to the issue of preservation of Weyl invariance,
there is therefore the issue of preservation of this additional gauge invariance.
This theory space thus offers an interesting opportunity to study
the RG flow in theory spaces admitting subspaces with special properties.
Somewhat similar issues appear in topologically massive gravity
and in three-dimensional higher derivative gravity.
The RG flows studied in \cite{tmg,tmsg} and \cite{ohta1,op}
did not preserve the special subspaces, in those cases.
In the case studied here we can construct flows that either preserve or do not preserve
the special subspace.

We close this section with an overview of the paper.
In section 2 we review the general formalism of Weyl gauging.
The quadratic expansion of the action is presented in section 3. 
We then add a cutoff term with a cutoff scale $k$. 
The crucial feature of the procedure is that this cutoff does not break Weyl invariance.
Following \cite{percacci,cdpp}, we parametrize the flow in terms of
the dimensionless, constant, Weyl invariant parameter $u=k/\chi$.
The beta functions are the derivatives of the couplings with respect to $u$.
In section 4 we give some details of their derivation and study fixed points.
These beta functions do not preserve the subspace where $b_\mu$ is massless.
In section 5 we discuss an alternative definition the preserves it.
Section 6 contains a brief summary.

\section{The classical action} \label{sec: The classical action}

The first step in the construction of a diffeomorphism-- and Weyl--invariant action
is the definition of the covariant derivative \eqref{wcovder} and curvature \eqref{wcurv}.
Let us note that both the covariant derivative $D_\mu$ and
the curvature tensor $\cR_{\mu\nu}{}^\rho{}_\sigma$ depend on the ``Weyl charge'' 
of the field, $w$.
If $w=0$ we have $\cR_{\mu\nu}{}^\rho{}_\sigma=\hat R_{\mu\nu}{}^\rho{}_\sigma$,
and we can further express $\hat R_{\mu\nu}{}^\rho{}_\sigma$ in terms of the
Riemann tensor $R_{\mu\nu}{}^\rho{}_\sigma$ (the curvature of the Levi-Civita connection) as
\bea
\label{wcovcurv}
\hat R_{\mu\nu\rho\sigma}&=& R_{\mu\nu\rho\sigma}
-F_{\mu\nu}g_{\rho\sigma}
+g_{\mu\rho}\left(\nabla_\nu b_\sigma+ b_\nu b_\sigma\right)
-g_{\mu\sigma}\left(\nabla_\nu b_\rho+ b_\nu b_\rho\right)
\nonumber
\\
&&
\!\!\!\!\!\!\!\!\!\!\!\!\!\!\!\!\!\!
-g_{\nu\rho}\left(\nabla_\mu b_\sigma+ b_\mu b_\sigma\right)
+g_{\nu\sigma}\left(\nabla_\mu b_\rho+ b_\mu b_\rho\right)
-\left(g_{\mu\rho}g_{\nu\sigma}-g_{\mu\sigma}g_{\nu\rho}\right)b^2\ ,
\eea
where $F_{\mu\nu}=\partial_\mu b_\nu-\partial_\nu b_\mu$ is the curvature
of the Weyl gauge field $b_\mu$.
Since $\hat\nabla$ is not metric, its curvature is not symmetric in the
second pair of indices:
\be
\hat R_{\mu\nu\rho\sigma}+\hat R_{\mu\nu\sigma\rho}=-2F_{\mu\nu}g_{\rho\sigma}\ .
\ee
There are thus two independent ``Ricci tensors'', obtained contracting the first
index of the curvature with the third or the fourth.
We will only need one of these definitions,
and we observe that the trace of this ``Ricci tensor'' is unique:
\bea
\label{curv}
\hat R_{\mu\nu}
\equiv \hat R_{\rho\mu}{}^\rho{}_\nu
&=&R_{\mu\nu}+F_{\mu\nu}+(d-2)(\nabla_\mu b_\nu+b_\mu b_\nu)
+\nabla^\rho b_\rho g_{\mu\nu}-(d-2)b^2g_{\mu\nu}\ ,
\\
\hat R&=&R+2(d-1)\nabla^\mu b_{\mu}-(d-1)(d-2)b^2\ .
\eea
The curvature of the connection $D_\mu$ acting on a vector of weight $w$ is
\be
\label{elizabeth}
\cR_{\mu\nu}{}^\rho{}_\sigma=\hat R_{\mu\nu}{}^\rho{}_\sigma
-w F_{\mu\nu}\,\delta^\rho_\sigma\ .
\ee

The simplest diffeomorphism-- and Weyl--invariant actions constructed only with the metric and $b_\mu$ are of the form
$c_1{\cal R}^2+c_2{\cal R}_{\mu \nu}{\cal R}^{\mu \nu}
+c_3{\cal R}_{\mu \nu \rho \sigma}{\cal R}^{\mu \nu \rho \sigma}
+c_4F_{\mu \nu}F^{\mu \nu}$.
We observe that changing the value of $w$, the first three terms generate
further contributions of the type of the fourth term.
In order to establish a basis of independent field monomials we thus have to fix the
value of $w$. In the following we will use $w=0$, which seems the most natural choice.
In this case the curvatures $\cR_{\mu\nu\rho\sigma}$ coincide with $\hat R_{\mu\nu\rho\sigma}$.

These actions contain also terms with four derivatives.
In addition to the metric and gauge field $b_\mu$
we will postulate the existence of a scalar $\chi$ with weight $w=-1$,
entirely analogous to the dilaton of \cite{percacci}.
Its covariant derivative is thus
\begin{equation}
D_\mu \chi =(\partial_\mu +b_\mu)\chi .
\end{equation}
If we restrict ourselves to actions that contain at most two derivatives of the fields,
we have the following four--parameter family of actions \cite{dirac}:
\begin{equation}
S=\int d^4x \sqrt{g} \left[
\frac{g_1}{2} D_\mu \chi D^\mu \chi 
+g_2\chi^4 
+\frac{g_3}{4} F_{\mu \nu} F^{\mu \nu} 
-g_4 \chi^2 {\cal R} \right]\ .
\label{action1}
\end{equation}
Every term in the above action is separately Weyl invariant. 
The equations of motion that follow from this action, 
written in explicitly Weyl--covariant form, are
\begin{eqnarray}
\label{eomg}
0&=&  - g_1D^2 \chi +4g_2\chi^3 -2g_4\chi {\cal R}  
\\
\label{eomb}
0&=& -g_3 D_\nu F^{\nu \mu} +(g_1+12 g_4) \chi D^\mu \chi 
\\
\label{eomchi}
0&=& g_4\chi^2\left(\cR^{\mu\nu}-\frac{1}{2}g^{\mu\nu}\cR\right)
-\frac{g_3}{2}\left(F^{\mu \rho}F^\nu{}_\rho
-\frac{1}{4}g^{\mu \nu}F_{\alpha\beta} F^{\alpha\beta}\right)
\nonumber 
\\
&&-\frac{g_1}{2}\left(D^\mu\chi D^\nu\chi
-\frac{1}{2}g^{\mu \nu}D_\rho\chi D^\rho\chi\right)
+\frac{1}{2}g_2 g^{\mu \nu}\chi^4
+g_4\left(g^{\mu \nu} D^2 \chi^2-D^{(\mu} D^{\nu )} \chi^2\right).
\end{eqnarray}

In \cite{percacci} the special case was studied when the Weyl connection is flat:
$F_{\mu \nu}=0$. This case can be obtained as follows.
With the dilaton one constructs a ``pure gauge'' Weyl vector
\be
\label{puregauge}
s_\mu=-\chi^{-1}\partial_\mu\chi\ .
\ee
One can use this gauge field to construct a covariant derivative $D^{(s)}$
and a curvature $\cR^{(s)}$, as in equations (\ref{wcovder},\ref{wcurv}).
When ambiguities can arise we will denote the previously defined covariant derivative and curvature
of $b_\mu$ by $D^{(b)}$ and $\cR^{(b)}$.
Note that
\be
\label{dschi}
D^{(s)}_\mu\chi=0\ .
\ee
In \cite{percacci} the integrable gauge field $s_\mu$ was used instead of $b_\mu$.
Note that at the classical level this can be seen as a special solution of the equations of motion:
from \eqref{eomb} one sees that if $g_1+12 g_4\not=0$, $F_{\mu\nu}=0$ implies
$D_\mu\chi=0$, which in turn is solved by $b_\mu = s_\mu$.
If we use this condition in the action, it reduces to:
\begin{equation}
\label{action3}
\int \sqrt{g} \left[g_2 \chi^4 
-g_4\chi^2\cR^{(s)}\right] 
= \int \sqrt{g} \left[g_2 \chi^4 
-g_4\chi^2 \left(R-6\chi^{-1}\nabla^2\chi\right) \right]\ .
\end{equation}
As already observed in \cite{deser}, the kinetic term of $\chi$ has the wrong sign
(note that here we are writing the Euclidean action).
This action is Weyl--invariant even without the Weyl gauge field.
It is said to be obtained from that
of a massless scalar by ``Ricci gauging'' \cite{iorio}.

This theory is just ordinary general relativity, with cosmological constant,
rewritten in Weyl--invariant form by use of a compensator field.
In fact, from the assumption that $\chi>0$ everywhere and from the transformation
property $\chi\to\Omega^{-1}\chi$ one deduces the existence of a gauge
where $\chi$ is constant. We can set
\be
\label{conversion}
g_4\chi^2=\frac{1}{16\pi G}\ ;\qquad
g_2\chi^2=2g_4\Lambda\ .
\ee
Then the action \eqref{action3} becomes just
\be
\label{hilbert}
S(g) = \frac{1}{16\pi G}\int d^4 x \sqrt{g}(2\Lambda-R)\ .
\ee

Now let us observe that
\be
\cR^{(b)}=\cR^{(s)}+6\chi^{-1}D^2\chi\ .
\ee
Using this and the rule for integration by parts (\ref{byparts})
one finds that (\ref{action1}) can be rewritten in the form:
\begin{equation}
S=\int d^4x \sqrt{g} \left[
\frac{g_1+12g_4}{2} D_\mu \chi D^\mu \chi 
+g_2\chi^4 
+\frac{g_3}{4} F_{\mu \nu} F^{\mu \nu} 
-g_4 \chi^2 \cR^{(s)} \right]\ .
\label{action2}
\end{equation}
This form makes it clear that a Higgs phenomenon is at work in this theory.
Going to the gauge (\ref{conversion}) the action reads
\begin{equation}
S(g) = \int d^4 x \sqrt{g}\left[\frac{1}{16\pi G}(2\Lambda-R)
+\frac{g_3}{4} F_{\mu \nu} F^{\mu \nu}
+\frac{g_1+12g_4}{32\pi G g_4} b_\mu b^\mu\right]\ ,
\label{higgsed}
\end{equation}
describing gravity coupled to a massive vector field.
In the special case when $g_1+12 g_4=0$, the Weyl gauge field is massless
and we are left with
\begin{equation}
\label{action4}
\int \sqrt{g} \left[\frac{g_3}{4}F_{\mu \nu}F^{\mu \nu}
+ g_2 \chi^4 
-g_4\left(\chi^2 R+6(\nabla\chi)^2\right)  \right]\ .
\end{equation}
This is the same as \eqref{action3}, plus the action of an abelian vector field
that is decoupled from $\chi$.
As a result, while the general action \eqref{action1} is only
invariant under the Weyl transformation
\be
g'_{\mu\nu}=\Omega^2 g_{\mu\nu}\ ,\qquad
b'_\mu=b_\mu+\Omega^{-1}\partial_\mu\Omega\ ,\qquad
\chi'=\Omega^{-1}\chi\ ,
\ee
the action \eqref{action4} is additionally invariant under the ``modified Weyl transformation''
where $b_\mu$ is inert:
\be
\label{modweyl}
g'_{\mu\nu}=\Omega^2 g_{\mu\nu}\ ,\qquad
b'_\mu=b_\mu\ ,\qquad
\chi'=\Omega^{-1}\chi\ ,
\ee
This additional invariance is a consequence of the fact that the Maxwell action
in four dimensions is invariant under Weyl transformations when the gauge field
is treated as a field of Weyl weight zero.
One can reparametrize these two gauge invariances as modified Weyl transformations
and ordinary abelian gauge transformations
\be
g'_{\mu\nu}=g_{\mu\nu}\ ,\qquad
b'_\mu=b_\mu+g^{-1}\partial_\mu g\ ,\qquad
\chi'=\chi\ ,
\ee
where $g$ is a gauge transformation parameter.
Thus \eqref{action4} can be interpreted as the action of conformal gravity
(or equivalently the action of a Ricci--gauged scalar)
coupled to an abelian gauge field
which has nothing to do with Weyl transformations.

In the following we will refer to the subspace defined by the equation
$g_1+12 g_4=0$ as the ``massless subspace''.
One of the main goals of this paper is to understand how
Weyl invariance can be maintained under quantization in the non-integrable Weyl theory
and in particular whether the massless subspace is preserved by the renormalization group flow.

\section{The quadratic action}

In this section we give the second variation of the action,
which is required for the quantization of the theory.
We will use the background field method.
For each field we choose generic background values,
henceforth denoted $g_{\mu\nu}$, $b_\mu$ and $\chi$ and expand:
\be
g_{\mu \nu}  \rightarrow g_{\mu \nu}+h_{\mu \nu}\ ;\qquad
b_{\mu}  \rightarrow  b_{\mu}+w_{\mu}\ ;\qquad
\chi  \rightarrow  \chi + \eta\ .
\ee
To second order in $h_{\mu\nu}$, $w_\mu$ and $\eta$, the action (\ref{action1}) becomes
\begin{eqnarray}
\frac{1}{2}&&\!\!\!\!\!\!\!\!\!
\int dx\,\sqrt{g}\, \Biggl\lbrace
g_4 \chi^2\,\bigg(-\frac{1}{2}h^{\mu\nu}D^2h_{\mu\nu}
+h^{\mu\nu}D_\mu D_\rho h^\rho{}_\nu
-h D_\mu D_\nu h^{\mu\nu}
+\frac{1}{2}h D^2 h
\nonumber \\
&+& \cR^{\mu\nu} h\,h_{\mu\nu}
-\cR^{\mu\nu}h_{\mu\rho}h_\nu{}^\rho
-{\cR}_{\alpha\mu\beta\nu}h^{\mu\nu}h^{\alpha\beta}\bigg) 
- g_4 D^\rho\chi^2
\left(
2h D_\sigma h^\sigma{}_\rho
+h_{\rho\nu}D_\sigma h^{\sigma\nu}\right)
\nonumber \\
&+& g_4 D^2\chi^2
\left(\frac{1}{4}h^2-\frac{3}{4}h^{\mu\nu}h_{\mu\nu}\right)%
+g_4 D^\mu D^\nu\chi^2 (h_\mu{}^\rho h_{\nu\rho}-2h\,h_{\mu\nu})
\nonumber \\
&+&\left[\frac{g_1}{2}(D\chi)^2+g_2\chi^4+\frac{g_3}{4}F^2-g_4\chi^2{\cal R}\right]
\left(\frac{1}{4}h^2-\frac{1}{2}h^{\mu\nu}h_{\mu\nu}\right)
\nonumber \\
&+&g_1 D_\mu\chi D_\nu\chi\left[h^{\mu\alpha}h^\nu_\alpha 
-\frac{1}{2}h h^{\mu\nu}\right] 
+\frac{g_3}{2}\left[F_{\mu\nu}F_\rho{}^\nu(2 h^{\mu\alpha}h^\rho{}_\alpha-h h^{\mu\rho}) 
+F_{\mu\nu}F_{\rho\sigma}h^{\mu\rho}h^{\nu\sigma} \right]
\nonumber\\
&+&g_3w_\mu
\left(-g^{\mu \nu} D^2+D^\mu D^\nu+{\cal R}^{\mu\nu}\right)w_\nu 
+(g_1+12g_4)\chi^2 w_\mu w^\mu
\nonumber\\
&+&\eta\left(-g_1 D^2+12g_2\chi^2-2g_4\cR\right)\eta
\nonumber \\
&+& g_3 F^{\mu\nu}h D_\mu w_\nu
-2g_3F_\rho{}^\nu h^{\mu \rho}\left(D_{\mu} w_{\nu}-D_{\nu} w_{\mu}\right)
+(g_1+12g_4)\chi D^\mu \chi
\left(h w_\mu-2h_{\mu\nu}w^\nu\right) 
\nonumber \\
&+&g_1D^\mu\chi\left(h  D_\mu\eta -2h_{\mu\nu} D^\nu\eta\right) 
+4g_4\chi\left(D^2h 
-D^\mu D^\nu h_{\mu\nu}
+{\cal R}_{\mu\nu} h^{\mu\nu}
-\frac{1}{2}{\cal R}h\right)\eta
\nonumber \\
&+&4g_2\chi^3 h\,\eta
-2(g_1+12g_4)\chi\eta D_{\mu} w^\mu
\Biggr\rbrace\ .
\end{eqnarray}
We have chosen to collect first the terms quadratic in $h$, $w$, $\eta$
and then the mixed terms $h$-$w$, $h$-$\eta$ and $w$-$\eta$.
The origin of each term can be easily traced by looking at the 
coefficients $g_1$, $g_2$, $g_3$ and $g_4$.

\subsection{The gauge fixing}

The quadratic action has zero modes corresponding to infinitesimal
diffeomorphisms $\xi$ and infinitesimal Weyl transformations $\omega$:
\bea
h_{\mu\nu}&=&\cL_\xi g_{\mu\nu}\ ;\qquad
w_\mu=\cL_\xi b_\mu\ ;\qquad
\eta=\cL_\xi \chi\ ,
\\
h_{\mu\nu}&=&2\omega g_{\mu\nu}\ ;\qquad
w_\mu=\partial_\mu\omega\ ;\qquad
\eta=-\omega\chi\ .
\eea
Quantization requires a nondegenerate operator,
which is achieved by adding a suitable gauge fixing condition.
In the background field method, the gauge fixing is designed
so as to preserve the ``background transformations''
\bea
\delta^{(D)}_\xi g_{\mu\nu}&=\cL_\xi g_{\mu\nu}\ ;\qquad
\delta^{(D)}_\xi b_\mu=\cL_\xi b_\mu\ ;\qquad
\delta^{(D)}_\xi\chi&=\cL_\xi \chi\ ;
\\
\delta^{(D)}_\xi h_{\mu\nu}&=\cL_\xi h_{\mu\nu}\ ;\qquad
\delta^{(D)}_\xi w_\mu=\cL_\xi w_\mu\ ;\qquad
\delta^{(D)}_\xi\eta&=\cL_\xi \eta\ ,
\\
\delta^{(W)}_\omega g_{\mu\nu}&=2\omega g_{\mu\nu}\ ;\qquad
\delta^{(W)}_\omega b_\mu=\partial_\mu\omega\ ;\qquad
\delta^{(W)}_\omega\chi&=-\omega\chi\ ,
\\
\delta^{(W)}_\omega h_{\mu\nu}&=2\omega h_{\mu\nu}\ ;\qquad
\delta^{(W)}_\omega w_\mu=0 \ ;\qquad
\delta^{(W)}_\omega\eta&=-\omega\eta\ ,
\eea
For the sake of defining a Weyl-covariant ghost operator it is convenient
to define modified diffeomorphism generators \cite{dhr}
\be
\tilde\delta^{(D)}_\xi =\delta^{(D)}_\xi+\delta^{(W)}_{-\xi^\mu b_\mu}\ .
\ee

A gauge fixing term for diffeomorphisms that manifestly preserves
the background gauge transformations is 
\be
S_{GF} =  \frac{g_4}{2\alpha}\int d^4x 
\sqrt{g}\,\chi^2 F_\mu {\bar g}^{\mu\nu} F_\nu\ ,
\ee
where
\be
\label{gf}
F_\nu =D_\mu h^\mu{}_\nu -\frac{1}{2}D_\nu h\ .
\ee
The ghost action corresponding to the gauge \eq{gf} is given by
\be
S_{gh}=-\int d^4x \sqrt{g}\, {\bar C}^\mu\left(\delta_\mu^\nu D^2
+\cR_\mu{}^\nu\right) C_\rho\ ,
\label{gh}
\ee
where $\bar C$ and $C$ are anticommuting vector fields.
To gauge--fix Weyl invariance we impose that $\eta=0$, a condition that does not lead to ghosts.
With this condition we can simply delete from the Hessian the rows and columns
that involve the $\eta$ field and we remain with a Hessian that is a quadratic
form in the space of the covariant symmetric tensors $h_{\mu\nu}$.

In the following we will choose the Feynman--de Donder gauge
$\alpha=1/g_4$, which simplifies the kinetic operators.
With these choices the gauge fixing can be expanded as
\begin{eqnarray}
S_{GF} &=&
\frac{g_4}{2} \int d^4 x\sqrt{g} 
\bigg[
\chi^2
\left(-h_{\mu \nu}D^\mu D_\rho h^{\rho \nu}
+hD^\mu D^\nu h_{\mu \nu}
-\frac{1}{4}hD^2 h\right)  
\nonumber
\\
&&\qquad\qquad+D_\mu\chi^2\left(
-h^{\mu \nu} D^\rho h_{\rho \nu} 
+h D_\nu h^{\mu\nu} 
-\frac{1}{4} h D^\mu h\right) \bigg]\ .
\end{eqnarray}
In \cite{percacci} the gauge fixing had the same form, but with $b_\mu$
replaced by $s_\mu$. Because of (\ref{dschi}), the second line vanished.

When the gauge fixing is taken into account, the total quadratic action
takes the form
\begin{eqnarray}
\frac{1}{2}&&\!\!\!\!\!\!\!\!\!
\int dx\,\sqrt{g}\, \Biggl\lbrace
g_4 \chi^2\,\bigg(-\frac{1}{2}h^{\mu\nu}D^2h_{\mu\nu}
+\frac{1}{2}h D^2 h
+\cR^{\mu\nu} h\,h_{\mu\nu}
-\cR^{\mu\nu}h_{\mu\rho}h_\nu{}^\rho
-{\cR}_{\alpha\mu\beta\nu}h^{\mu\nu}h^{\alpha\beta}\bigg) 
\nonumber \\
&-&  g_4 D^\rho\chi^2
\left(h D_\sigma h^\sigma{}_\rho
+2h_{\rho\nu}D_\sigma h^{\sigma\nu}\right)
\nonumber \\
&+& g_4 D^2\chi^2
\left(\frac{3}{8}h^2-\frac{3}{4}h^{\mu\nu}h_{\mu\nu}\right)
+g_4 D^\mu D^\nu\chi^2 (h_\mu{}^\rho h_{\nu\rho}-2h\,h_{\mu\nu})
\nonumber \\
&+& 
\left[\frac{g_1}{2}(D\chi)^2+g_2\chi^4+\frac{g_3}{4}F^2-g_4\chi^2{\cal R}\right]
\left(\frac{1}{4}h^2-\frac{1}{2}h^{\mu\nu}h_{\mu\nu}\right)
\label{totalvar}
\\
&+&g_1D_\mu\chi D_\nu\chi \left[h^{\mu\alpha}h^\nu_\alpha 
-\frac{1}{2}h h^{\mu\nu}\right]
+\frac{g_3}{2}\left[F_{\mu\nu}F_\rho{}^\nu
(2 h^{\mu\alpha}h^\rho{}_\alpha-h h^{\mu\rho}) 
+F_{\mu\nu}F_{\rho\sigma} h^{\mu\rho}h^{\nu\sigma}
\right]
\nonumber
\\
&+&g_3w_\mu
\left(-g^{\mu\nu} D^2+D^\mu D^\nu+{\cal R}^{\mu\nu}\right)w_\nu 
+(g_1+12g_4)\chi^2 w_\mu w^\mu
\nonumber 
\\
&+& g_3 F^{\mu \nu}hD_\mu w_\nu
-2g_3F_\rho{}^\nu h^{\mu \rho}\left(D_{\mu} w_{\nu}-D_{\nu} w_{\mu}\right)
+(g_1+12g_4)\chi D^\mu \chi
\left(h w_\mu-2h_{\mu\nu}w^\nu\right) 
\Biggr\rbrace\ .
\nonumber
\end{eqnarray}
Note in particular that the last term in the second last line
is a mass term for $w_\mu$ proportional to $g_1+12g_4$,
in accordance with the previous statement that a Higgs phenomenon
is occurring in this theory.

For technical reasons it proves convenient to decompose the field $w_\mu$ 
into its transverse and longitudinal components.
We refer to appendix \ref{appendix york decomposition} for some details.

\subsection{The cutoff}

We use the formalism of the effective average action, which is an effective action $\Gamma_k$ calculated in the presence of an infrared cutoff of the form
\be
\label{cutoff}
\Delta S_k=\frac{1}{2}\int dx\sqrt{g} \Psi \cR_k(-D^2)\Psi\ .
\ee
Here $\Psi=(h_{\mu\nu},w_\mu)$ is the multiplet formed by the fluctuation fields
and $\cR$ is a matrix in field space containing the couplings $g_i$,
times a cutoff profile function $R_k$ which we choose to be
$R_k(z)=(k^2-z)\theta(k^2-z)$ \cite{optimized}.
We neglect the $k$-derivatives of the couplings in the cutoff (``one loop approximation'').

The $k$-derivative of the effective average action satisfies the Wetterich equation \cite{wetterichEq}
\be
k\frac{d\Gamma_k}{dk}=\frac{1}{2} \Tr\left(
\frac{\delta^2\Gamma_k}{\delta\Psi\delta\Psi}+\cR_k\right)^{-1}k\frac{d\cR_k}{dk}\ .
\ee
The r.h.s. of this equation is the ``beta functional'' of the theory,
the generating functional of all beta functions (in the sense that
the coefficient of some field monomial is the beta function of the corresponding coupling).
Of course the effective average action will generally contain infinitely many terms,
including non--local ones,
but here we concentrate our attention just on the ones of the form appearing 
in the action \eqref{action1}.

An elegant way of calculating the beta functions of the four couplings $g_1$, $g_2$, $g_3$, $g_4$
is to insert an ansatz for $\Gamma_k$ of the form \eqref{action1}
in the Wetterich equation and to extract the coefficient of the relevant field monomials.
The resulting beta functions are obviously not exact, since other terms are generated in the flow
and their contribution is neglected. Nevertheless this procedure can yield valuable information
and has been used in many studies of the gravitational beta functions.
The main issue we are interested in here is the preservation of Weyl invariance along the flow.
The question has been answered previously in the case of ``dilaton gravity'', 
namely when the Weyl gauge field is flat \cite{percacci,cdpp}.
Here we will discuss the case of a non--flat Weyl gauge field,
and the special case of Ricci--gauging, $g_1+12g_4=0$.

In order to maintain Weyl invariance along the flow,
the computation is carried out along the same lines as in \cite{percacci}.
The following procedure is used.
First, as already indicated in \eqref{cutoff}, the cutoff is chosen to be a function of
the Weyl--covariant operator $-D^2$.
Then, instead of thinking of $\log k$ as the independent variable of the flow,
we assume that the cutoff $k$ is proportional to $\chi$ and we
take the Weyl--invariant, dimensionless, constant quantity $u=k/\chi$ as independent variable.
Thus, the couplings will be functions of $u$.
The cutoff can be rewritten
\be
\Delta S_k(-D^2)=\frac{1}{2}\int dx\sqrt{g}\,\chi^2\Psi (u^2-\cO)\theta(u^2-\cO) \Psi
\ee
where $\cO=-(1/\chi^2)D^2$.

Since the r.h.s. of the Wetterich equation is the trace of a function 
of a Weyl--covariant operator, it is Weyl--invariant.
Using heat kernel methods, it can be expanded as a sum of monomials
constructed with the background fields, their derivatives, and the curvatures.
By isolating terms of the form \eqref{action1} one reads the
beta functions of the couplings $g_i$.

\section{The RG flow}

\subsection{Derivation of the beta functions}

In order to project out the beta functions of the various couplings 
one has to calculate some terms of the functional trace on the r.h.s. of the Wetterich equation.
In order to simplify the calculations,
we take advantage of the independence of the results from the choice of background,
and we choose for each coupling/beta function the simplest background
that makes the corresponding field monomial nonzero.
As long as the same gauge condition and cutoff are used in all calculations,
the result is the same as computing the functional trace with a general background.

To calculate the beta function of $g_1$ we choose backgrounds with
$\cR_{\mu\nu\rho\sigma}=0$, $F_{\mu\nu}=0$ but $D_\mu\chi\not=0$.
To extract the terms proportional to $(D\chi)^2$, we note that the full
kinetic operator $\cO=\frac{\delta^2\Gamma_k}{\delta\Psi\delta\Psi}$, 
which can be read off \eqref{totalvar},
can be expanded as
\begin{equation}
\cO=\cO_0+P^{(1)}+P^{(2)} \label{expansion1}
\end{equation}
where $\cO_0=\cO|_{D\chi\rightarrow 0}$,
$P^{(1)}$ are the terms of order $D\chi^2$ and 
$P^{(2)}$ are the terms of order $(D\chi)^2$ or $D^2\chi^2$. 
We treat $P^{(1)}$ and $P^{(2)}$ as perturbations and expand
\begin{equation}
\frac{1}{\cO+\cR_k(-D^2)}=G-G P^{(1)}G-G P^{(2)}G+G P^{(1)}G P^{(1)}G+\ldots  \label{expansion2}
\end{equation}
where $G=\frac{1}{\cO_0+\cR_k(-D^2)}$.
One then has to evaluate a trace of a function of $-D^2$
with some insertions of powers of $D_\mu$.
Such traces can be evaluated using the ``universal RG machine'' 
developed in \cite{RgMachine,OffDiagonalHk}.

To calculate the beta function of $g_3$ we need a background with $F_{\mu\nu}\not=0$.
Since $F_{\mu\nu}$ appears in the kinetic operator, we can proceed as in the preceding case,
assuming that $F^2$ is small and expanding as in (\ref{expansion1},\ref{expansion2}).
The $h$-$h$ part of the second variation contains a term of order $F^2$ while the
non-diagonal terms are of order $F$.
There are thus contributions linear in $P^{(2)}$ and quadratic in $P^{(1)}$, 
multiplied by the the heat kernel coefficient $B_0(-D^2)$.
There is another potential source of $F^2$ terms:
it consists of terms of order zero in the perturbations
proportional to the heat kernel coefficient $B_4(-D^2)$.
Indeed the latter contains terms quadratic in curvature, which themselves contain $F$.
As noted earlier, these terms depend on the choice of basis of invariant operators.
Furthermore, we do not currently have the formula for the $b_4$ coefficient of $-D^2$
(in Appendix D we have evaluated only the coefficient $b_2$).
For this reason we shall leave this contribution in the form of an undetermined
coefficient $K$ in the beta function of $g_3$
(see equation (\ref{beta13}) below).
We observe that this contribution is easily distinguishable from the remaining ones,
which are proportional to $g_3$, whereas the one coming from $b_4$ is purely numerical.

As manifested in \eqref{conversion},
the couplings $g_2$ and $g_4$ are related to the cosmological constant and 
Newton's constant respectively, so to calculate their beta functions
one needs some curved background.
The simplest possibility is to choose the background $b_\mu=0$, $\chi$ constant,
since in this way our quadratic action reduces to the usual linearized Einstein-Hilbert action 
plus a minimally coupled massive vector field $w_\mu$.
The calculation is similar to the one in \cite{percacci},
except for the presence of the Weyl vector.
Of course in this way one does not see explicitly that Weyl invariance is preserved
by the beta functions: one has to appeal to the Weyl invariance of the general construction.
As an additional check, in Appendix D we show that this particular choice is not necessary
and that Weyl invariance emerges explicitly.

The flow equation for the E-H terms is:
\begin{eqnarray}
\partial_t \Gamma_k &=& \frac{1}{(4\pi)^2} \int d^4x \sqrt{g} 
\Bigg\lbrace 5 Q_2\left(\frac{\partial_t R_k}{P_k-\frac{g_2}{g_4} \chi^2} \right)
-4Q_2\left(\frac{\partial_t R_k}{P_k}\right) 
+\cR\left[\frac{5}{6}Q_1\left(\frac{\partial_t R_k}{P_k-\frac{g_2}{g_4} \chi^2} \right)
 \right. 
\nonumber \\
&-& \left. 
\frac{2}{3} Q_1\left(\frac{\partial_t R_k}{P_k}\right)
-3 Q_2\left(\frac{\partial_t R_k}{(P_k-\frac{g_2}{g_4}\chi^2)^2}\right)
-Q_2\left(\frac{\partial_t R_k}{P_k^2}\right)\right]
\label{ehflow}
\\
&+&
\frac{3}{2} 
Q_2\left(\!\frac{\partial_t R_k}{P_k+\frac{g_1+12 g_4}{g_3}\chi^2}\!\right) 
+R \left[
\frac{1}{8}Q_1\left(\!\frac{\partial_t R_k}{P_k+\frac{g_1+12 g_4}{g_3}\chi^2}\!\right)
-\frac{3}{8} Q_2\left(\!\frac{\partial_t R_k}{\left(P_k+\frac{g_1+12 g_4}{g_3}\chi^2\right)^2}\!\right)
\right]   
\Bigg\rbrace 
\nonumber
\end{eqnarray}
The first two lines contain the contribution of the graviton and ghost,
see e.g. equation (39) in \cite{cpr2}.
The last line gives the contribution of the transverse part of the gauge field $w_\mu$. 
The contribution of the longitudinal part of $w_\mu$
is cancelled by that of the scalar field which takes the jacobian 
of the decomposition into account.

Finally we collect here all the beta functions.
Denoting $\beta_i=u\frac{dg_i}{du}$, we find
\begin{eqnarray}
\beta_1 &=&  \frac{1}{16 \pi ^2} 
 \left[ 
\frac{3(g_1+12g_4)^2}{g_3g_4\left(1-\frac{g_2}{g_4 u^2}\right)^2\left(1+\frac{g_1+12 g_4}{g_3 u^2}\right)}
+\frac{3(g_1+12g_4)^2}{g_3g_4\left(1-\frac{g_2}{g_4 u^2}\right)\left(1+\frac{g_1+12 g_4}{g_3 u^2}\right)^2}
-8 u^2
\label{beta11}
\right. 
\\
&& \left. 
\!\!\!\!\!\!\!\!\!\!\!\!\!\!\!\!\!\!\!\!\!
-\frac{2 u^2 (g_1+12g_4)}{3 g_4 \left(1-\frac{g_2}{g_4 u^2}\right)}
-\frac{2 u^2 (g_1+12g_4)}{3 g_4 \left(1-\frac{g_2}{g_4 u^2}\right)^2}
 +\frac{12 u^2}{\left(1-\frac{g_2}{g_4 u^2}\right)^3}
 +\frac{3 (g_1+12 g_4)}{g_3 \left(1+\frac{g_1+12 g_4}{g_3 u^2}\right)}
 +\frac{3 (g_1+12 g_4)}{g_3 \left(1+\frac{g_1+12 g_4}{g_3 u^2}\right)^2}
 \right] 
\nonumber 
\\
\beta_2 &=& \frac{u^4}{16 \pi ^2}   
\left[-4+\frac{3}{2 \left(1+\frac{g_1+12 g_4}{g_3u^2}\right)}
+\frac{5}{1-\frac{g_2}{g_4 u^2}}\right]   
\label{beta12}
\\
\beta_3 &=&  \frac{1}{16 \pi ^2} \left[\frac{K}{1-\frac{g_2}{g_4 u^2}}
-\frac{3g_3  u^2}{g_4 \left(1-\frac{g_2}{g_4  u^2}\right)^2} 
+\frac{2 g_3 u^2}{g_4}
\frac{\left(1-\frac{g_2}{g_4u^2}\right)+\left(1+\frac{g_1+12 g_4}{g_3 u^2}\right)}{\left(1-\frac{g_2}{g_4u^2}\right)^2\left(1+\frac{g_1+12 g_4}{g_3 u^2}\right)^2}
\right] 
\label{beta13}
\\
\beta_4 &=& \frac{u^2}{16 \pi ^2} \left[\frac{7}{3}
-\frac{1}{4 \left(1+\frac{g_1+12 g_4}{g_3 u^2}\right)}
+\frac{3}{8 \left(1+\frac{g_1+12 g_4}{g_3 u^2}\right)^2}
-\frac{5}{3 \left(1-\frac{g_2}{g_4 u^2}\right)}
+\frac{3}{\left(1-\frac{g_2}{g_4 u^2}\right)^2}\right]
\label{beta14}
\end{eqnarray}

\subsection{Fixed points}

In the standard Wilsonian approach to the renormalization group
one uses the cutoff $k$ as independent variable and also measures 
all dimensionful couplings in units of $k$.
This leads to flow equations that are autonomous, meaning that
the independent variable does not appear explicitly,
but only as argument of the running couplings.
In this context the definition of fixed point is very simple:
it is just a zero of the beta functions.
To find the fixed points one need not solve the flow equations,
which are differential equations:
it is enough to solve a system of algebraic equations.

The price we have to pay for manifest Weyl--invariance
is that the beta functions contain $u$ explicitly: the flow is not autonomous.
In this situation it is generally not obvious how fixed points can be defined,
since any zeroes of the beta functions will in general move
as functions of the renormalization group time $t=\log u$.
It looks like any analysis of the flow will require solving differential equations.
Fortunately one can again reduce the flow to autonomous equations:
if one performs the redefinitions
\be
\label{redef}
g_1=f_1 u^2\ ;\qquad 
g_2=f_2 u^4\ ;\qquad
g_3=f_3 \ ;\qquad 
g_4=f_4 u^2\ ,
\ee
in the beta functions $u$ factors, leaving only overall powers
that can be cancelled between the left and right hand sides of the flow equations.
Then one can find fixed points for $f_1$, $f_2$, $f_3$ and $f_4$ in the usual way.
\footnote{Note that $g_1$ can be eliminated from the action by a redefinition of $\chi$
and would be expected to be redundant, {\it i.e.} to disappear from the beta functions.
Why this does not happen has been explained in \cite{perini3}.
See also \cite{dietzmorris2}.}

It is easy to see why this procedure should work.
From \eqref{action1}, note that the powers of $u$ in \eqref{redef} 
are equal to the power of $\chi$ in the corresponding field monomial.
Also recall that it is possible to go to the gauge where $\chi$ is constant.
Then one can absorb the powers of $\chi$ in the coupling constants and
the powers of $\chi$ are the mass dimensions of these dimensionful couplings.
But then one sees that the couplings $f_i$ are just the usual Wilsonian couplings
made dimensionless by dividing them by powers of $k$,
and we know that such couplings satisfy autonomous flow equations.

Solving numerically the fixed point equations for the couplings $f_i$,
one finds several real solutions.
Recalling that $g_3$ can be seen as the inverse of the QED coupling $e^2$,
one expects a fixed point at $e^2=0$, which obviously is not
visible in the original parameterization.
If we rewrite the RG equations in terms of $e^2$ one indeed finds a fixed point at
\begin{center}
\begin{tabular}{ | l | r | r | r | r | r | r |}
\hline
  & $f_{1*}$   & $f_{2*}$ & $e^2_*$ & $f_{4*}$ & $\tilde\Lambda_*$ & $\tilde G_*$ \\
\hline
FP$_1$ & $\phantom{-}0.0161$  & $\phantom{-}0.008585$ & $\phantom{-}0.0000$ &  $0.02327$ & $0.1845$  & $0.8549$ \\ \hline
\end{tabular}
\end{center}
For the sake of comparison with the literature we have given here also the fixed point values of
$$
\tilde\Lambda=\frac{f_2}{2f_4}\\ ;\qquad
\tilde G=\frac{1}{16\pi f_4}\ .
$$
(These relations follow from (\ref{conversion}) and (\ref{redef}).)
The following table gives the eigenvalues of the linearized flow,
ordered from the most to the least relevant
\begin{center}
\begin{tabular}{ | l | r | r | r | r |}
\hline
  & $\lambda_1$   & $\lambda_2$ & $\lambda_3$ & $\lambda_4$  \\
\hline
FP$_1$ & $-2.74294$ & $-2.28003 + 1.96824 i$ & $-2.28003 - 1.96824 i$ & $-0.179136$ \\ \hline
\end{tabular}
\end{center}
The corresponding eigenvectors are $f_1$, complex mixtures of $f_1$, $f_2$, $f_4$ and
a mixture mostly along $f_3$, respectively.
There is also the true Gaussian fixed point with $G=0$,
which would require a further change of variable.
The properties of these two fixed points are independent 
of the value of the undetermined constant $K$.
In addition there are three real fixed points with $e^2\not=0$, whose properties
depend to some extent on $K$. This dependence is not very strong, however,
and we have checked that their qualitative properties would 
be the same for a wide range of values of $K$.
\begin{center}
\begin{tabular}{ | l | r | r | r | r | r | r |}
\hline
  & $f_{1*}$   & $f_{2*}$ & $f_{3*}$ & $f_{4*}$ & $\tilde\Lambda_*$ & $\tilde G_*$ \\
\hline
FP$_2$ & $0.47503$   & $0.004118$ & $0.0000$ & $0.01698$ & $0.1213$ & $1.1718$\\ \hline
FP$_3$ & $0.0382$ & $0.008357$ & $4.8524$ & $0.02291$ & $0.1823$ & $0.8681$ \\ \hline
FP$_4$ & $-0.1493$  & $-0.005328$ & $-0.0488$ &  $0.01531$  & $-0.1493$ & $1.5387$ \\ \hline
\end{tabular}
\end{center}
and the eigenvalues of the linearized flow
\begin{center}
\begin{tabular}{ | l | r | r | r | r |}
\hline
  & $\lambda_1$   & $\lambda_2$ & $\lambda_3$ & $\lambda_4$  \\
\hline
FP$_2$ & $-1.95041$ & $-1.86782+1.39828i$ & $-1.86782-1.39828i$ & $-0.811311$\\ \hline
FP$_3$ & $-2.36814$ & $-2.25983+1.99667i$ & $-2.25983+1.99667i$ & $0.175235$\\ \hline
FP$_4$ & $-2.78268$ & $-2.12422$ & $-1.54627$ & $15.3501$ \\ \hline
\end{tabular}
\end{center}
We do not list the eigenvectors but we note the following: 
at FP$_2$ there is a clean separation between the eigenvector of $\lambda_1$,
which is a mixture of $g_1$ and $g_3$,
the eigenvector of $\lambda_4$ which is exactly $g_3$
and the complex eigenvalues, which have no component on $g_3$;
at FP$_3$ the eigenvectors have a very similar structure, but they all
have some component on all couplings;
at FP$_4$ all eigenvectors have a strong component only along $g_1$.
This, together with the unphysical values of $g_1$ and $g_3$
make this an uninteresting, probably spurious fixed point and we shall not consider it further.

We have explored the properties of these fixed points for $-10<K<10$.
All the listed parameters of FP$_2$ and FP$_3$ change only on the second or third significant 
digit for $K$ in this range.
The value of $f_{3*}$ for FP$_2$ has the same sign as $K$ and ranges between $\pm0.04$.

From these tables, knowing that in the gauge where $\chi$ is constant the theory reduces to
Einstein-Hilbert gravity coupled to a massive vector field, and comparing with
results from the literature, one may venture to say that FP$_1$ and one between FP$_2$ and FP$_3$
probably correspond to known fixed points and may have some physical relevance
whereas the other two are most likely artifacts of the truncation.

\section{An alternative cutoff}

Now we consider in greater detail the subspace of theory space where the Weyl field is massless.
As we have seen in section \ref{sec: The classical action}, if we set $g_1+12 g_4=0$ 
we recover the Weyl integrable theory with a massless, minimally coupled abelian gauge field. 
The action reduces to the form (\ref{action4}), which, aside from the presence 
of the abelian gauge field, has been discussed in detail in \cite{percacci,cdpp}. 
It is natural to ask whether the RG flow preserves this subspace.
To this effect, one has to compute the beta function of $g_1+12 g_4$ and check whether 
it is zero when one sets $g_1+12g_4=0$. From (\ref{beta11},\ref{beta14}) one sees that this is not the case.

The reason for this is not hard to understand.
The massless subspace $g_1+12g_4=0$ is characterized by the enlarged symmetry (\ref{modweyl}).
The operator $-D^{{(b)}2}$ which was used in the definition of the cutoff
is not covariant under the transformations (\ref{modweyl}), where $b_\mu$ is inert.
Thus the beta functions do not preserve the enlarged symmetry.
This immediately suggests an alternative cutoff procedure:
to define the cutoff using the operator $-D^{{(s)}2}$, 
which, being independent of $b_\mu$, is covariant
both under ordinary and modified Weyl transformations.
In this section we discuss the calculation of the beta functions obtained from this alternative regularization procedure. 

\subsection{The modified beta functions}

The calculation of the beta functions of $g_2$ and $g_4$,
with the background $b_\mu=0$ and $\chi$ constant, is exactly as in section 4.
Thus $\beta_2$ and $\beta_4$ remain as in (\ref{beta12},\ref{beta14}).
The calculation of the beta function of $g_1$ also proceeds along the same lines as before
but now there are some differences: in the second variation of the action (\ref{action2})
the terms containing derivatives of $\chi$ (second and third line in (\ref{totalvar}))
are now zero because of (\ref{dschi}).
This removes several contributions to $\beta_1$.
In the case of $g_3$ the term proportional to $b_4(-D^2)$ is now absent,
because now $D$ is $D^{(s)}$ and the fields strength of $s_\mu$ is zero.
Thus if we choose a basis for operators containing powers of 
$\hat R^{(s)}_{\mu\nu\rho\sigma}$ (namely the curvature given in
equation (\ref{curv}), with $b_\mu$ replaced by $s_\mu$), 
there is no contribution to $\beta_3$ from $b_4(-D^2)$,
in other words we can set the parameter $K=0$.

With these modifications, one arrives at the following beta functions:
\begin{eqnarray}
\beta_1 &=&  \frac{1}{16 \pi ^2} 
 \left[-28u^2
+\frac{3(g_1+12 g_4)^2}{g_3g_4\left(1-\frac{g_2}{g_4 u^2}\right)^2 
\left(1+\frac{g_1+12 g_4}{g_3 u^2}\right)}
+\frac{3(g_1+12 g_4)^2}{g_3g_4\left(1-\frac{g_2}{g_4 u^2}\right) 
\left(1+\frac{g_1+12 g_4}{g_3 u^2}\right)^2}
\right. 
\nonumber 
\\
&& \left. 
-\frac{2 u^2 (g_1-18g_4)}{3 g_4 \left(1-\frac{g_2}{g_4 u^2}\right)}
-\frac{2 u^2 (g_1+66g_4)}{3 g_4 \left(1-\frac{g_2}{g_4 u^2}\right)^2}
+\frac{3 u^2}{\left(1+\frac{g_1+12 g_4}{g_3 u^2}\right)}
-\frac{9u^2}{2\left(1+\frac{g_1+12 g_4}{g_3 u^2}\right)^2}
\right] 
\label{beta21}
\\
\beta_3 &=& 
\!\!\!\!
\frac{u^2}{16 \pi ^2}\frac{g_3}{g_4} 
\left[
\frac{2}{\left(1-\frac{g_2}{g_4 u^2}\right)^2 \left(1+\frac{g_1+12 g_4}{g_3 u^2}\right)}
+\frac{2}{\left(1-\frac{g_2}{g_4 u^2}\right) \left(1+\frac{g_1+12 g_4}{g_3 u^2}\right)^2}
-\frac{3}{\left(1-\frac{g_2}{g_4 u^2}\right)^2}
\right]\ ,
\label{beta23}
\end{eqnarray}
while the other two have remained as in (\ref{beta12},\ref{beta14}).
We see that that for $g_1+12g_4=0$,
$\beta_1+12\beta_4=0$, so the massless subspace is indeed invariant.

\goodbreak

\subsection{Redundancy and fixed points}

Let us consider the action, written in the form (\ref{action2}),
choose the gauge where $\chi$ is constant and use equations (\ref{redef}) and the relation
$u=k/\chi$ to write
\begin{equation}
S=\int d^4x \sqrt{g} \left[
\frac{f_1+12f_4}{2} k^2b^2 
+f_2 k^4 
+\frac{f_3}{4} F_{\mu \nu} F^{\mu \nu} 
-f_4 k^2 \cR^{(s)} \right]\ .
\label{action5}
\end{equation}
In this gauge $f_3$ can be seen as the coefficient of the kinetic term for the vector while 
the mass is given by the combination $f_1 + 12 f_4$. It is clear that via a suitable
rescaling of $b$ one can eliminate either $f_3$ or $f_1+12f_4$.
If we redefine the couplings as
\be
\label{redef2}
f_1+12f_4=Z_b \ess_1\ ;\qquad 
f_2=\ess_2\ ;\qquad
f_3=Z_b  \ ;\qquad 
f_4=\ess_4\ ,
\ee
then $Z_b$ can be eliminated by a redefinition of $b_\mu$:
it is a redundant coupling. 
\footnote{This would no longer be true if $b_\mu$ was coupled to some matter field.}
(We consider the alternative choice in appendix E.)
This interpretation is confirmed by the explicit form of the beta functions:
\bea
\beta_{\ess_1}&=& -2\ess_1 +\eta_b \ess_1 
+\frac{(1-2\ess_1)(2-\ess_1)\ess_2-(4-10\ess_1-5\ess_1^2)\ess_4}
{48 \pi ^2 \left(\ess_1+1\right){}^2 \left(\ess_2-\ess_4\right){}^2}\ess_1
\nonumber\\
\beta_{\ess_2}&=& -4 \ess_2 -\frac{8 \ess_2 \ess_1 +5 \ess_2 + 2 \ess_4 \ess_1 + 5 \ess_4}{32 \pi ^2 (1+\ess_1 ) (\ess_2-\ess_4)}
\\
\beta_{\ess_4}&=&  -2\ess_4 + 
\frac{(56\ess_1^2+106\ess_1 +59)\ess_2^2
-6(12\ess_1^2+22\ess_1+13)\ess_2\ess_4
+(88\ess_1^2+170\ess_1+91)\ess_4^2}{384 \pi ^2 (1+\ess_1)^2 (\ess_2-\ess_4)^2}
\nonumber
\eea
and the anomalous dimension
\be
\eta_b= -\frac{\beta_Z}{Z}=\frac{2 \ess_2 +\ess_4 \left(3 \ess_1^2+4 \ess_1 -1\right)}{16 \pi ^2 (1+\ess_1)^2 (\ess_2-\ess_4)^2} \ ,
\ee
which only depend on the essential couplings $\ess_i$.

The system of three equations $\beta_{\ess_i}=0$ admits three real fixed points
with $\ess_1$ finite or zero, and one with $1/\ess_1=0$:
\begin{center}
\begin{tabular}{ | l | r | r | r | r | r | r |}
\hline
  & $\eta_{b*}$   & $1/\ess_{1*}$ & $\ess_{2*}$ & $\ess_{4*}$ & $\tilde\Lambda_*$ & $\tilde G_*$ \\ \hline
FP$_2$ & $1.9504$  & $0$ & $0.00411798$ & $0.0169775$ & $0.1213$ & $1.1718$ \\ \hline
\hline
  & $\eta_{b*}$   & $\ess_{1*}$ & $\ess_{2*}$ & $\ess_{4*}$ & $\tilde\Lambda_*$ & $\tilde G_*$ \\ \hline
FP$_1$ & $-0.179136$  & $0$ & $0.00858496$ & $0.0232715$ & $0.184452$ & $0.854881$ \\ \hline
FP$_3$ & $1.27047$  & $1.28633$ & $0.00628253$ & $0.0198675$ & $0.158111$ & $1.00135$\\ \hline
FP$_4$ & $1.76077$  & $-2.5551$ & $0.000150384$ &  $0.0127745$ & $0.00588611$ 
& $1.55735$ \\ \hline
\end{tabular}
\end{center}
The inverted numbering of the first two fixed points is deliberate:
it is such that the values of $\ess_2$ and $\ess_4$ are equal to
the values of $f_2$ and $f_4$ for the fixed point by the same name in section 4.2.
This suggests that perhaps these fixed points can be identified.
This observation is strengthened by the results for the eigenvalues:
\begin{center}
\begin{tabular}{ | l | r | r | r |}
\hline
  & $\lambda_1$   & $\lambda_2$ & $\lambda_3$  \\
\hline
FP$_2$ & $-1.86782+1.39828 i$ & $-1.86782-1.39828 i$ & $-1.1391$ \\ \hline
FP$_1$ & $-2.92208$ & $-2.28003 + 1.96824 i$ & $-2.28003 - 1.96824 i$ \\ \hline
FP$_3$ & $-2.02559 + 1.87941 i$ & $-2.02559 - 1.87941 i$ & $0.923836$\\ \hline
FP$_4$ & $-3.13639$ & $-1.40315$ & $3.36778$ \\ \hline
\end{tabular}
\end{center}
The eigenvectors at FP$_2$ are complex mixtures of $\ess_2$ and $\ess_4$,
and a mixture mostly along $1/\ess_1$, respectively.
The eigenvectors at FP$_1$ are complex mixtures of $\ess_2$ and $\ess_4$,
and a mixture mostly along $\ess_1$, respectively.
It is interesting to note that FP$_1$ lies in the massless subspace,
since $\ess_1=0$ means $g_1+12 g_4=0$. 
The linearized flow tells us that this choice is attractive in the UV.

We observe that the eigenvalues $\lambda_2$ and $\lambda_3$ of FP$_1$ coincide
with $\lambda_2$ and $\lambda_3$ of the fixed point FP$_1$ in section 4.2.
Furthermore, the anomalous dimension is equal to $-\lambda_1$ of FP$_1$ in section 4.2.
Similar identifications can be made for FP$_2$, suggesting that these four
fixed points can be identified pairwise. (This was the motivation for the names
in the first place.)
The identification of FP$_3$ and FP$_4$ with the other two fixed points of
section 4.2 is also relatively obvious, but in these two cases
the values do not coincide numerically.

\subsection{Inside the massless subspace}

In the preceding section we have considered a set of beta functions in the full theory space
that preserve the massless subspace.
The correct beta functions inside the massless subspace are however different,
since they must take into account the enlarged gauge symmetry that is present there.
As already noted, in the massless subspace the theory is equivalent to gravity coupled 
to a Maxwell field. 
Therefore, one has to add a gauge fixing and a ghost term for the new abelian gauge symmetry
(the abelian ghost is decoupled in flat space but it contributes to the beta functions
of $g_2$ and $g_4$ because it is coupled to gravity).
There is no need to add these terms outside the massless subspace.
Here we discuss the modifications that follow.

We choose a standard Lorentz gauge condition, such that the gauge fixing and ghost terms are
\be
S_{\rm gf}+S_{\rm gh}=\int d^4x\left[\frac{1}{2\alpha}(D_\mu w^\mu)^2
+\frac{1}{\sqrt{\alpha}}\bar c\nabla^2 c\right].
\ee
These have to be added to the quadratic action.
As in the preceding sections, we decompose the field $w^\mu$ into its transverse
and longitudinal components $w_\mu=w_\mu^T+D_\mu(-D^2)^{-1/2}\tilde\phi$.
This transformation has a trivial Jacobian so the new terms in the action amount to
\be
S_{\rm gf}+S_{\rm gh}=\int d^4x\left[\frac{1}{2\alpha}\tilde\phi(-D^2)\tilde\phi
+\frac{1}{\sqrt{\alpha}}\bar c(-D^2) c\right]
\ee
which contributes to the beta functional
\be
\frac{1}{2}\Tr\left[\frac{\partial_tR_k(-D^2)}{-D^2+R_k(-D^2)}\right]
-\Tr\left[\frac{\partial_tR_k(-D^2)}{-D^2+R_k(-D^2)}\right]\ .
\ee
The additional contribution is therefore equivalent to that of an anticommuting real scalar.
\footnote{If we had chosen the gauge $\alpha=0$, which amounts to imposing the gauge condition strongly, one would not have the contributions from $\tilde\phi$ and the ghosts but instead
there would be the contribution from the Jacobian of the decomposition,
which is again equivalent to an anticommuting scalar.}

The calculation of the beta functions of $g_2$ and $g_4$,
if we choose the background $b_\mu=0$ and $\chi$ constant, is exactly the
same as in section 4.
This only changes the contribution of the abelian vector field
given in the last line of equation (\ref{ehflow}), which now becomes
\be
\frac{1}{(4\pi)^2}\int d^4x\sqrt{g}
\left\lbrace
Q_2\left(\!\frac{\partial_t R_k}{P_k}\!\right)
+R \left[
\frac{1}{24}Q_1\left(\!\frac{\partial_t R_k}{P_k}\!\right)
-\frac{3}{8} Q_2\left(\!\frac{\partial_t R_k}{P_k^2}\!\right)
\right]\right\rbrace\ .
\ee
It is interesting to compare this to the result given in equation (23) of \cite{cpr2}:
\be
\frac{1}{(4\pi)^2}\int d^4x\sqrt{g}
\left\lbrace
Q_2\left(\!\frac{\partial_t R_k}{P_k}\!\right)
+R \left[
\frac{1}{6}Q_1\left(\!\frac{\partial_t R_k}{P_k}\!\right)
-\frac{1}{2} Q_2\left(\!\frac{\partial_t R_k}{P_k^2}\!\right)
\right]\right\rbrace\ .
\ee
In both cases one is using a cutoff ``of type I'', but the difference lies in the fact
that here we decompose the vector field into its transverse and longitudinal parts,
and impose cutoffs separately, whereas in \cite{cpr2} no such decomposition was used.
Numerically, when one uses the optimized cutoff,
the coefficient of $R$ turns out to be $-7/24$ in the first case and $-4/24$ in the second.

The new terms bring only small changes to the beta functions of section 5.1:
the beta function $\beta_1$ is as in (\ref{beta21}) except that $-28$ is replaced by $-30$;
the beta function $\beta_2$ is as in (\ref{beta12}) except that $-4$ is replaced by $-9/2$;
the beta function $\beta_3$ remains as in (\ref{beta23});
the beta function $\beta_4$ is as in (\ref{beta14}) except that $7/3$ is replaced by $5/2$.
The beta functions written in this way are extensions of ones valid in the massless 
subspace to the whole theory space.
The flow they describe is very similar to the one described in section 5.1,
aside from minor numerical corrections which are anyway within the
theorietical uncertainties of this type of calculation.
There is however no reason to gauge fix outside the massless subspace,
so these beta functions are strictly speaking not correct there.
They are correct in the massless subspace $g_1+12g_4=0$,
where they reduce to the following simple beta functions:
\begin{eqnarray}
\beta_2 &=& \frac{u^4}{16 \pi ^2}   
\left[-3+\frac{5}{1-\frac{g_2}{g_4 u^2}}\right]   
\label{beta31}
\\
\beta_3 &=&  \frac{u^2}{16 \pi ^2}\frac{g_3}{g_4}
\left[\frac{2}{\left(1-\frac{g_2}{g_4 u^2}\right)}
-\frac{1}{\left(1-\frac{g_2}{g_4 u^2}\right)^2}
\right] 
\label{beta32} \\
\beta_4 &=& \frac{u^2}{16 \pi ^2} \left[\frac{21}{8}
-\frac{5}{3 \left(1-\frac{g_2}{g_4 u^2}\right)}
+\frac{3}{\left(1-\frac{g_2}{g_4 u^2}\right)^2}\right] \label{beta33}
\end{eqnarray}
We have not written $\beta_1$ since it is still true that $\beta_1=-12\beta_4$.
Furthermore, note that $g_1$ does not appear in any of the other beta functions at all.

There are now only two fixed points: one with $g_3=0$ and one with $1/g_3\equiv e^2=0$.
We list here their properties:
\begin{center}
\begin{tabular}{ | l | r | r | r | r | r |}
\hline
  &  $f_{2*}$ & $f_{3*}$ & $f_{4*}$ & $\tilde\Lambda_*$ & $\tilde G_*$ \\
\hline
FP$_2$   & $0.007013$ & $0.0000$ & $0.0214442$ & $0.163518$ & $0.927726$\\ \hline
  &  $f_{2*}$ & $e^2_*$ & $f_{4*}$ & $\tilde\Lambda_*$ & $\tilde G_*$ \\
\hline
FP$_1$  & $0.007013$ & $0.0000$ & $0.0214442$ & $0.163518$ & $0.927726$\\ \hline
\end{tabular}
\end{center}
The eigenvalues are as follows:
\begin{center}
\begin{tabular}{ | l | r | r | r |}
\hline
  & $\lambda_1$   & $\lambda_2$ & $\lambda_3$  \\
\hline
FP$_2$ & $-2.14278 + 1.75252 i$ & $-2.14278-1.75252 i$ & $0.225565$\\ \hline
FP$_1$ & $-2.14278 + 1.75252 i$ & $-2.14278-1.75252 i$ & $-0.225565$ \\ \hline
\end{tabular}
\end{center}
with the complex eigenvalues referring to a mixture of $g_2$ and $g_4$,
while the real eigenvalue is for $g_3$.

If one neglects the threshold effects represented by the nontrivial denominators,
and uses the definitions (\ref{conversion}), (\ref{beta32}) becomes just
\be
\beta_3=\frac{1}{\pi}g_3\tilde G\ .
\label{beta43}
\ee
Without the coupling to gravity the field $b_\mu$ would be just a free
vector field and its beta function would vanish.
The beta function (\ref{beta43}) is entirely due to the effect of the
gravitational coupling. This effect has been the subject of some
interest in recent years \cite{rw,pietrykowski,toms,epr,flp}.
One should not attach to these beta functions the same physical
meaning of the usual perturbative beta functions \cite{anberetal}.
The calculation we have done here is very similar to the one
in \cite{harst} and finds a nonvanishing, positive coefficient.
We note that if $b_\mu$ was coupled to some charged fields,
for example as in QED, there would be an additional constant
contribution $-C$ to (\ref{beta43}).
This would then translate into a beta function for $e^2$ of the form
\be
Ce^4-\frac{1}{\pi}\tilde G e^2\ ,
\ee
which is indeed of the form found in \cite{harst}.
If $C>0$, as is the case in QED, this, together with the beta functions 
for $\tilde\Lambda$ and $\tilde G$ admits, 
in addition to FP$_1$ and FP$_2$ also a third fixed point with finite, nonzero $e^2$
and $e^2$ irrelevant.

\section{Concluding remarks}

This is the third and last paper in a series that deals with the definition of the
RG flow in Weyl-invariant theories. In \cite{percacci} it was established that
in Weyl gravity (\ref{action2}) one can define the cutoff in such a way
that the flow is explicitly Weyl-invariant.
In \cite{cdpp} the result was extended to theories including matter,
both minimally coupled and self-interacting.
In both papers the Weyl-covariant derivative was constructed with the
pure gauge vector potential $s_\mu$ of (\ref{puregauge}).
In this paper we have discussed the case when the Weyl vector potential
$b_\mu$ is non-integrable.

The four-parameter class of actions (\ref{action1}) defines an interesting system that
exhibits Weyl invariance and, in a codimension-one subspace where $b_\mu$
is massless, an additional abelian gauge invariance.
The main result of the preceding papers, namely the existence of a Weyl-invariant flow,
has been reestablished in the full theory space.
In addition we have examined here the peculiarities arising from the existence of
the massless subspace.
If one defines the cutoff using the Laplacian $-D^{(b)2}$ the flow does not
leave the massless subspace invariant.
In spite of the possibility of rescaling away one of the parameters of the action,
none of the couplings is redundant and the RG flow admits some zeroes
for the system of all four beta functions.
If one defines the cutoff using the Laplacian $-D^{(s)2}$ the flow leaves 
the massless subspace invariant.
One of the parameters is redundant and the remaining system of three 
beta functions has some fixed points roughly corresponding to those of
the preceding procedure.
The massless subspace is UV attractive.
Finally, the RG flow for the couplings inside the massless subspace
must take into account the gauge fixing and ghost terms associated to
the enlarged abelian gauge invariance and are therefore slightly different
from those outside the subspace.

In the calculation of the beta functions for the couplings $g_2$ and $g_4$
(related to the cosmological constant and Newton's contant)
we have used a maximally symmetric background metric, with
background $b_\mu=0$ and $\chi$ constant.
This is sufficient to determine the beta functions but Weyl invariance of the
flow is not manifest.
We have shown in Appendix D that the relevant heat kernel coefficients are actually Weyl invariant.
The calculation could thus have been done on an arbitrary background.
This answers a minor issue that has remained lingering for some time.
The beta function of Newton's coupling had been computed in \cite{reuterweyer}
in the so-called ``CREH'' approximation, where only the conformal factor
of the metric is dynamical.
In this approximation the Einstein-Hilbert action has the form (\ref{action3}),
where $R$ is the curvature of a fixed reference metric.
One can then read the beta function of Newton's coupling (or equivalently the
``anomalous dimension'' $\eta=\partial_tG/G$) either from the second
or from the third term of (\ref{action3}).
Since the second term can be viewed as a potential for the conformal factor $\chi$
the result was denoted $\eta^{(\rm pot)}$, and since the third term is the ordinary
kinetic term of $\chi$ the result was called $\eta^{(\rm kin)}$.
The two calculations in \cite{reuterweyer} gave $\eta^{(\rm kin)}\not=\eta^{(\rm pot)}$,
so the question remained whether a quantization exists for which 
$\eta^{(\rm kin)}=\eta^{(\rm pot)}$.
We have shown that the answer is positive.


\appendix

\section{Weyl covariant derivatives acting on $h_{\mu \nu}$: a list}

We report here the explicit expression of some terms with Weyl covariant derivatives
and curvatures that enter in the second variation of the action. 
\begin{eqnarray}
h^{\mu \nu} D^2 h_{\mu \nu} &=& h^{\mu \nu}\nabla^2 h_{\mu \nu} +4h^{\mu \nu}b_{\mu} \nabla^\rho h_{\rho \nu} -4h^{\mu \nu} b^\rho \nabla_{\mu} h_{\rho \nu}-2h^{\mu \nu} b^\rho \nabla_\rho h_{\mu \nu}-8h^{\mu \nu} b_\mu b^\rho h_{\rho \nu}-2h^{\mu \nu} h_{\mu \nu} b^2 \nonumber \\
&+& 4hh^{\mu \nu}b_\mu b_\nu \nonumber \\
h_{\mu \nu}D^\mu D_\rho h^{\rho \nu} &=&h_{\mu \nu} \left[ \nabla^\mu \nabla_\rho h^{\rho \nu} -4\nabla^\mu b_\rho \cdot h^{\rho \nu} -4b_\rho \nabla^\mu h^{\rho \nu} +\nabla^\mu b^\nu \cdot h +b^\nu \nabla^\mu h-g^{\mu \nu} b_\alpha \nabla_\beta h^{\alpha \beta} \right.  \nonumber \\
&+& \left. \left( -g^{\mu \nu}b^2+2b^\mu b^\nu \right) h+ 4g^{\mu \nu} b_\alpha b_\beta h^{\alpha \beta} +2b^\mu \nabla_\rho h^{\rho \nu}-8b^\mu b_\rho h^{\rho \nu}   \right] \nonumber \\
\left( hD^\mu D^\nu h_{\mu \nu} \right) &=& h\left[ \nabla^\mu \nabla^\nu h_{\mu \nu}+\nabla^\mu b_\mu \cdot h +b^\mu \nabla_\mu h -4\nabla^\mu b^\nu \cdot h_{\mu \nu} -6b^\mu \nabla^\nu h_{\mu \nu} -2b^2 h+8b^\mu b^\nu h_{\mu \nu}  \right] \nonumber \\
hD^2 h &=& h\left[ \nabla^2 -2b^\mu \nabla_\mu  \right] h
\end{eqnarray}
and:
\begin{eqnarray}
h^{\mu \nu}h^{\alpha}_\mu {\cal R}_{\nu \alpha} &=& h^{\mu \nu}h^{\alpha}_\mu \left\lbrace R_{\nu \alpha} +2\nabla_\nu b_\alpha +g_{\nu \alpha} \nabla^{\sigma}b_\sigma +2b_\nu b_\alpha -2g_{\nu \alpha}b^2  \right\rbrace \nonumber \\
h^{\mu \nu}h^{\alpha \beta}{\cal R}_{\alpha \mu \beta \nu}&=&h^{\mu \nu}h^{\rho \sigma} \left\lbrace R_{\rho \mu \sigma \nu} +g_{\rho \sigma}(\nabla_\mu b_\nu +b_\mu b_\nu)-g_{\rho \nu}(\nabla_\mu b_\sigma +b_\mu b_\sigma)-g_{\mu \sigma}(\nabla_\rho b_\nu +b_\rho b_\nu) \right.  \nonumber \\
&+&\left.  g_{\mu \nu}(\nabla_\rho b_\sigma +b_\rho b_\sigma)-(g_{\rho \sigma}g_{\mu \nu}-g_{\rho \nu}g_{\mu \sigma})b^2 \right\rbrace \nonumber \\
hh^{\mu \nu}{\cal R}_{\mu \nu}&=& hh^{\mu \nu} \left\lbrace R_{\mu \nu} +2\nabla_\mu b_\nu +g_{\mu \nu} \nabla^{\sigma}b_\sigma +2b_\mu b_\nu -2g_{\mu \nu}b^2  \right\rbrace 
\end{eqnarray}
Finally:
\begin{eqnarray}
-h_{\mu \nu}D^\mu D_\rho h^{\rho \nu} &=&(-1)h_{\mu \nu} \left[ \nabla^\mu \nabla_\rho h^{\rho \nu} -4\nabla^\mu b_\rho \cdot h^{\rho \nu} -4b_\rho \nabla^\mu h^{\rho \nu} +\nabla^\mu b^\nu \cdot h +b^\nu \nabla^\mu h-g^{\mu \nu} b_\alpha \nabla_\beta h^{\alpha \beta} \right.  \nonumber \\
&+& \left. \left( -g^{\mu \nu}b^2+2b^\mu b^\nu \right) h+ 4g^{\mu \nu} b_\alpha b_\beta h^{\alpha \beta} +2b^\mu \nabla_\rho h^{\rho \nu}-8b^\mu b_\rho h^{\rho \nu}   \right] \nonumber \\
\frac{1+\beta}{4} \left( hD^\mu D^\nu h_{\mu \nu} \right) &=& \frac{1+\beta}{4} h\left[ \nabla^\mu \nabla^\nu h_{\mu \nu}+\nabla^\mu b_\mu \cdot h +b^\mu \nabla_\mu h -4\nabla^\mu b^\nu \cdot h_{\mu \nu} -6b^\mu \nabla^\nu h_{\mu \nu} -2b^2 h+8b^\mu b^\nu h_{\mu \nu}  \right] \nonumber \\
\frac{1+\beta}{4} \left( h_{\mu \nu} D^\mu D^\nu h  \right) &=& \frac{1+\beta}{4} \left[ h_{\mu \nu} \nabla^{\mu} \nabla^{\nu} h-hb^{\mu}\nabla_\mu h +h_{\mu \nu}\left(b^{\mu} \nabla^\nu + b^{\nu} \nabla^\mu \right) h \right]  \nonumber \\
-\frac{(1+\beta)^2}{16}hD^2 h &=& -\frac{(1+\beta)^2}{16} h\left[ \nabla^2 -2b^\mu \nabla_\mu  \right] h
\end{eqnarray}

\section{Weyl covariant decomposition}  \label{appendix york decomposition}

Let us recall that when a vector $A_\mu$ in a functional integral
is decomposed into its transverse and
longitudinal parts $A_\mu^T+\nabla_\mu\phi$, there arises a Jacobian
~\cite{mottola}:
\begin{equation}
1=\int {\cal D} A_\mu e^{-\frac{1}{2}\int dx \sqrt{g} A^\mu A_\mu}=J \int  {\cal D} A^{(T)}_\mu {\cal D} \phi e^{-\frac{1}{2}\int dx \sqrt{g} A^{(T)\mu} A^{(T)}_\mu+\phi (-\nabla^2) \phi}=J\int{\cal D} \phi e^{-\frac{1}{2}\int dx \sqrt{g} \phi (-\nabla^2) \phi}.
\end{equation}
The above gaussian normalized measure is not Weyl invariant. Since we want to use 
a Weyl invariant measure, the above steps are modified as follows.
Let $A_\mu=A^{(T)}_\mu +D_\mu \phi$ with $D^\mu A^{(T)}_\mu =0$, then:
\begin{equation}
1=\int {\cal D} A_\mu e^{-\frac{1}{2}\int dx \sqrt{g} \chi^2 A^\mu A_\mu}=J \int  {\cal D} A^{(T)}_\mu {\cal D} \phi e^{-\frac{1}{2}\int dx \sqrt{g} \chi^2  A^{(T)\mu} A^{(T)}_\mu+\phi (-D^2) \phi}=J\int{\cal D} \phi e^{-\frac{1}{2}\int dx \sqrt{g} \chi^2 \phi (-D^2) \phi}.
\end{equation}
Note that the above derivation hold if the background is such that $D\chi=0$. If this is not the case we have:
\begin{eqnarray}
1&=& \int {\cal D} A_\mu e^{-\frac{1}{2}\int dx \sqrt{g} \chi^2 A^\mu A_\mu}=J \int  {\cal D} A^{(T)}_\mu {\cal D} \phi e^{-\frac{1}{2}\int dx \sqrt{g} \chi^2  \left[ A^{(T)\mu} A^{(T)}_\mu +\phi (-D^2) \phi -\frac{D^\mu \chi^2}{\chi^2} A^{(T)}_{\mu} \phi -\phi \frac{D^\mu \chi^2}{\chi^2} D^\mu \phi \right]} \nonumber \\
J^{-1}&=& \int {\cal D} A_\mu^{(T)} {\cal D} \phi e^{-\frac{1}{2} \int \sqrt{g} \psi M \psi}
\end{eqnarray}
where $\psi=(A^{(T)}_\mu ,\phi)$ and $M$ is the following matrix:
\begin{equation}
M=\frac{\chi^2}{2} \left( \begin{array}{cc} g^{\mu}_\nu & - \frac{D_{\nu}\chi^2}{\chi^2} \\ 
- \frac{D^{\mu}\chi^2}{\chi^2} & -D^2- \frac{D^{\rho}\chi^2}{\chi^2} D_\rho \\ \end{array} \right)
\end{equation}

Since the above field are bosonic we need to evaluate $\det(M)$ which can be done introducing two auxiliary (grassmaniann odd) fields ($\xi_\mu, \tau$). In order to be able to exponentiate in the action this determinant we also perform the redefinition $\xi_\mu \rightarrow \frac{\sqrt{-D^2}}{\chi} \xi_\mu$. This further redefinition gives a jacobian which also has to be taken into account via another auxiliary field ($v_\mu$). 

\section{Rule for integration by parts}

We discuss here the integration by parts with the Weyl-covariant derivative (\ref{wcovder}).
As an illustration it will be sufficient to consider an integral of the form
\be
\int d^4x\sqrt{g}\,A D_\mu B^\mu\ ,
\ee
where $A$ is a scalar and $B$ is a vector. The case when $A$ and $B$ have additional
contracted indices works in the same way.
The important assumption that we have to make is that the integral is not only
invariant under diffeomorphisms, as is already clear, but also under Weyl transformations.
This is equivalent to saying that it is dimensionless.
Not all integrals need to be dimensionless, but this is the case for the action,
and this is the only case we are interested in in this paper.
Since $\sqrt{g}$ has weight $4$, if $A$ has weight $w_A$, $B$ must have $-4-w_A$.
Assuming that surface terms can be discarded we then find
\bea
\int d^4x\sqrt{g}\,A D_\mu B^\mu
&=&
\int d^4x\sqrt{g}\,A (\partial_\mu B^\mu+\hat\Gamma_\mu{}^\mu{}_\rho B^\rho+(4+w_A)b_\rho B^\rho)
\nonumber\\
&=&
\int d^4x\sqrt{g}\left[-B^\mu\frac{1}{\sqrt{g}}\partial_\mu(\sqrt{g}A)+A\Gamma_\mu{}^\mu{}_\rho B^\rho+w_A Ab_\rho B^\rho)\right]
\nonumber\\
&=&\int d^4x\sqrt{g}\left[-B^\mu\partial_\mu A+w_AAb_\rho B^\rho\right]
\nonumber\\
&=&-\int d^4x\sqrt{g}B^\mu D_\mu A\ .
\label{byparts}
\eea
We see that Weyl covariant derivatives can be integrated by parts
provided the integral is dimensionless.

\section{Some results on the heat kernel}

In the body of the paper we have calculated the beta functions using simple
backgrounds that are just sufficient to make the relevant invariants nonzero,
for example a spherical metric with $b_\mu=0$ and $\chi$ constant.
Gauge invariance was taken from the general construction of the RG flow
and was not checked explicitly.
Here we point out that gauge invariance follows from properties of the heat kernel of $-D^2$.
More precisely we have
\be
\label{hk}
b_0=\frac{1}{16\pi^2}\tr\mathrm{1}\ ;\qquad
b_2=\frac{1}{16\pi^2}\frac{1}{6}\tr\cR\mathbf{1}\ .
\ee
Since the connection $\hat{\Gamma}$ is non-metric we cannot apply directly known results, so
we express $-D^2$ in terms of $\nabla$, the Levi-Civita connection, and $b_\mu$.
For a scalar we can use the known property that for an operator 
$\Delta = -\nabla^2 + A^\mu \nabla_\mu +X$ 
the coefficients $b_2$ reads \cite{ggr}:
\begin{equation}
b_2(\Delta)= \frac{1}{(4 \pi)^{d/2}} \left[ \frac{R}{6}-X+\frac{1}{2} \nabla_\mu A^\mu -\frac{1}{4} A_\mu A^\mu \right]  \nonumber
\end{equation}
For a scalar of weight $-1$:
\begin{eqnarray}
-D^2 &=& -\nabla^2 \phi + b^\mu b_\mu \phi -\nabla^\mu b_\mu \cdot \phi .
\end{eqnarray}
Inserting in the above equation one obtains (\ref{hk}).
For the graviton the situation is more complicated 
since $-D^2$ contains terms which are of the form $\bi^\mu \nabla_\alpha h_{\mu \beta}$. 
To overcome this problem we expand the non-minimal terms in $e^{-s(-D^2)}$ and employ the off-diagonal HK coefficients \cite{RgMachine,OffDiagonalHk}. 
In this way one arrives again at (\ref{hk}).

\section{Alternative definition of the essential couplings}

In section 5.3 the redefinition
\be
\label{redef2}
g_1+12g_4=Z_b u^2\ ;\qquad 
g_2=\ess_2 u^4\ ;\qquad
g_3=Z_b \ess_3 \ ;\qquad 
g_4=\ess_4 u^2\ ,
\ee
provides an alternative division of the couplings into redundant and essential ones.
In this parametrization the beta functions are:
\bea
\beta_{\ess_2}&=&\frac{8\ess_2+2\ess_4+5(\ess_2+\ess_4)\ess_3}{32\pi^2(1+\ess_3)(\ess_4-\ess_2)}-4\ess_2
\nonumber\\
\beta_{\ess_3}&=&-\frac{\ess_3\left(3\ess_4+4\ess_3\ess_4+\ess_3^2(2\ess_2-\ess_4)\right)}{16\pi^2(1+\ess_3)^2(\ess_4-\ess_2)^2}+\eta_b\ess_3
\\
\beta_{\ess_4}&=&\frac{8(7\ess_2^2-9\ess_2\ess_4+11\ess_4^2)+2\ess_3(53\ess_2^2-66\ess_2\ess_4+85\ess_4^2)
+\ess_3^2(59\ess_2^2-78\ess_2\ess_4+91\ess_4^2)}{384\pi^2(1+\ess_3)^2(\ess_4-\ess_2)^2}-2\ess_4
\nonumber
\eea
and the anomalous dimension is
\be
\eta_b=-\frac{\beta_Z}{Z}=2-\frac{2\ess_2+5\ess_4
-5\ess_3(\ess_2-2\ess_4)+2\ess_3^2(\ess_2-2\ess_4)}{48\pi^2(1+\ess_3)^2(\ess_4-\ess_2)^2}\ .
\ee
The system of three equations $\beta_{\ess_i}=0$ admits four real fixed points,
one of which occurs at $\ess_3=\infty$ or $e^2=0$:
\begin{center}
\begin{tabular}{ | l | r | r | r | r | r | r |}
\hline
  & $\eta_{b*}$   & $\ess_{2*}$ & $e^2_*$ & $\ess_{4*}$ & $\tilde\Lambda_*$ & $\tilde G_*$ \\
\hline
FP$_1$ & $2.7272$  & $\phantom{-}0.008585$ & $\phantom{-}0.0000$ &  $0.02327$ & $\phantom{-}0.1237$ 
& $1.1338$ \\ \hline
\hline
  & $\eta_{b*}$   & $\ess_{2*}$ & $\ess_{3*}$ & $\ess_{4*}$ & $\tilde\Lambda_*$ & $\tilde G_*$ \\
\hline
FP$_2$ & $0.8113$  & $0.004118$ & $0.0000$ & $0.01698$ & $0.1213$ & $1.1718$\\ \hline
FP$_3$ & $1.2705$  & $0.006282$ & $0.7774$ & $0.01987$ & $0.1581$ & $1.0013$ \\ \hline
FP$_4$ & $1.7608$  & $0.000150$ & $-0.3914$ &  $0.01277$  & $0.0059$ & $1.5573$ \\ \hline
\end{tabular}
\end{center}
Note that the values of $\ess_2$ and $\ess_4$ at FP$_1$ and FP$_2$
are the same at the fixed point by the same name in section 5.2.
This identification is reinforced by the results for the eigenvalues:
\begin{center}
\begin{tabular}{ | l | r | r | r |}
\hline
  & $\lambda_1$   & $\lambda_2$ & $\lambda_3$  \\
\hline
FP$_1$ & $-2.92208$ & $-2.28003+1.96824 i$ & $-2.28003-1.96824 i$ \\ \hline
FP$_2$ & $-1.86782+1.39828 i$ & $-1.86782-1.39828 i$ & $-1.1391$\\ \hline
FP$_3$ & $-2.02559+1.87941 i$ & $-2.02559-1.87941 i$ & $0.923836$ \\ \hline
FP$_4$ & $-3.13639$ & $-1.40315$ & $3.36778$ \\ \hline
\end{tabular}
\end{center}
At FP$_1$ the complex eigenvectors are mixture of $\ess_2$, $\ess_4$
and the real, least relevant, eigenvalue is almost entirely $\ess_3$.

Note that the values of $\ess_2$, $\ess_3$ and $\ess_4$
and the eigenvalues $\lambda_2$ and $\lambda_3$
of FP$_1$ agree with those of the fixed point FP$_1$
of section 4.2.
Furthermore we note that the eigenvalues and the anomalous dimensions
of all fixed points are identical to those we found in the parameterization of section 5.2.
These are essentially the same fixed point described in different parameterizations.

\end{document}